\documentclass[%
reprint,
amsmath,amssymb,superscriptaddress,
aps,pra
]{revtex4-2}

\usepackage[table]{xcolor}

\usepackage{siunitx}
\usepackage{graphicx}
\usepackage{dcolumn}
\usepackage{svg}
\usepackage{bm}
\usepackage[unicode=true, colorlinks=true, citecolor={blue!80!black}, urlcolor={blue!50!black}, linkcolor = {blue!80!black}]{hyperref}
\usepackage{easyReview}
\usepackage{adjustbox}
\usepackage{xcolor}
\usepackage[T1]{fontenc}
\usepackage{blindtext}
\usepackage{braket}
\usepackage{gensymb}
\usepackage{booktabs}
\usepackage[euler]{textgreek}

\begin{document}
\preprint{QuTech/AndersenLab}

\title{Flip-chip integrated superconducting qubits using electroplated bump bonds}
\author{Yen-An Shih}
\author{Rebecca Gharibaan}
\author{Barka Khan}
\author{Dhananjay Joshi}
\author{Siddharth Singh}
\author{Martijn F. S. Zwanenburg}
\author{Eugene Y. Huang}
\author{Nataliia Zhurbina}
\author{Figen Yilmaz}
\author{Lukas Johannes Splitthoff}
\author{Srijit Goswami}
\author{Christian Kraglund Andersen}

\affiliation{QuTech and Kavli Institute for Nanoscience, Delft University of Technology, Delft 2628 CJ, The Netherlands}

\date{\today}

\begin{abstract}
Flip-chip integration offers a promising route toward scalable superconducting quantum processors and hybrid semiconductor-superconductor quantum devices. We develop a three-dimensional transmon architecture using electroplated indium in which the qubit electric field is shared nearly equally between two bump-bonded substrates while maintaining low participation at the indium-bump interface. The resulting geometry is well suited for future hybrid qubits, enabling the integration of distinct material platforms while minimizing sensitivity to bump-interface loss. Using this platform, we evaluate electroplated indium interconnects for superconducting quantum circuits. Flip-chip transmons incorporating electroplated indium bumps exhibit qubit quality factors around $10^6$. In addition, a systematic study of coplanar-waveguide resonators is used to identify losses associated with the electroplating process. In particular, we find that surface losses associated with the gold-layer, used to enable good electric contact with the indium, is likely the primary contributor to the qubit decay rate. These results demonstrate the compatibility of electroplated indium technology with high-coherence superconducting circuits and establish a promising platform for three-dimensional hybrid quantum integration.
\end{abstract}

\maketitle

\section{Introduction}

Superconducting quantum circuits are a leading platform for quantum information processing, with continued improvements in coherence times, gate fidelities, and system integration enabling increasingly complex quantum processors~\cite{Acharya2025, CarreraVazquez2024, Arute2019, Blais2020}. As the number of qubits increases, however, the wiring, packaging, and layout constraints of planar circuit architectures become increasingly challenging~\cite{Kosen_2022, Rosenberg2017, Blais2020,Brecht2016,Norris2024}. These challenges are particularly relevant for hybrid semiconductor-superconductor devices, where high-mobility semiconductor heterostructures~\cite{Scappucci2021, Wan2015}, gate electrodes, superconducting qubits, and microwave circuitry often impose competing fabrication and materials constraints~\cite{depalma2025lowlossfrequencytunablejosephsonjunction, Yu2023}. Flip-chip integration provides a route to separate these functionalities across multiple substrates, while maintaining compact microwave coupling and vertical interconnects between chips~\cite{Foxen_2018, Rosenberg2017, 10.1063/5.0050173,granel20263dintegrationhybridquantum}. More generally, by fabricating circuit elements on separate chips and subsequently bonding them together, one can separate the functions of the quantum device across multiple substrates, for example by placing qubits on one chip and control, readout, or filtering elements on another~\cite{Kosen_2022, PRXQuantum.5.030350}. This approach can improve routing flexibility, reduce the footprint of individual components, and enable more scalable device layouts. In addition, the flip-chip geometry provides a means to engineer the electromagnetic environment of superconducting circuits by distributing the electric field across substrates and interfaces~\cite{10.1063/5.0068255}. Such control over the electric-field participation ratio is particularly attractive for hybrid semiconductor-superconductor systems~\cite{Yu2023, PhysRevResearch.6.023094, aelm.202400687, PhysRevApplied.19.054026, PRXQuantum.5.030337}, since it may reduce sensitivity to the integrated, and comparatively lossy, semiconductor materials.

A central requirement for a flip chip architecture is a reliable interconnect between chips that provides both mechanical stability and low-resistance superconducting electrical contact. Indium bumps are widely used for flip-chip bonding because indium is mechanically soft, can accommodate surface nonuniformities during bonding, and becomes superconducting at millikelvin temperatures~\cite{Foxen_2018,10.1063/5.0003907}. In superconducting quantum circuits, indium bumps are commonly fabricated by evaporation followed by lift-off~\cite{Foxen_2018,10.1063/5.0003907,Kosen_2022,Rosenberg2017}. While this approach is compatible with many established cleanroom processes, it becomes increasingly challenging when thick bumps or large bump volumes are required. Beyond standard superconducting devices, large chip-separation is particularly attractive for hybrid semiconductor-superconductor devices~\cite{granel20263dintegrationhybridquantum, Holman2021}. Evaporating several micrometers of indium can require long deposition times and may lead to poor sidewall coverage and difficult lift-off~\cite{JIANG2004143, s25010263}. Electroplating provides an attractive alternative for high bumps, as it enables faster deposition of thick indium structures and may offer improved scalability to larger bump arrays and wafer-level processing~\cite{JIANG2004143, s25010263}. However, electroplated bumps introduce additional materials and interfaces, including seed layers and electrochemical processing steps, whose compatibility with high-coherence superconducting circuits must be carefully assessed.

\begin{figure*}[t]
    \centering
    \includegraphics[width=1\textwidth]{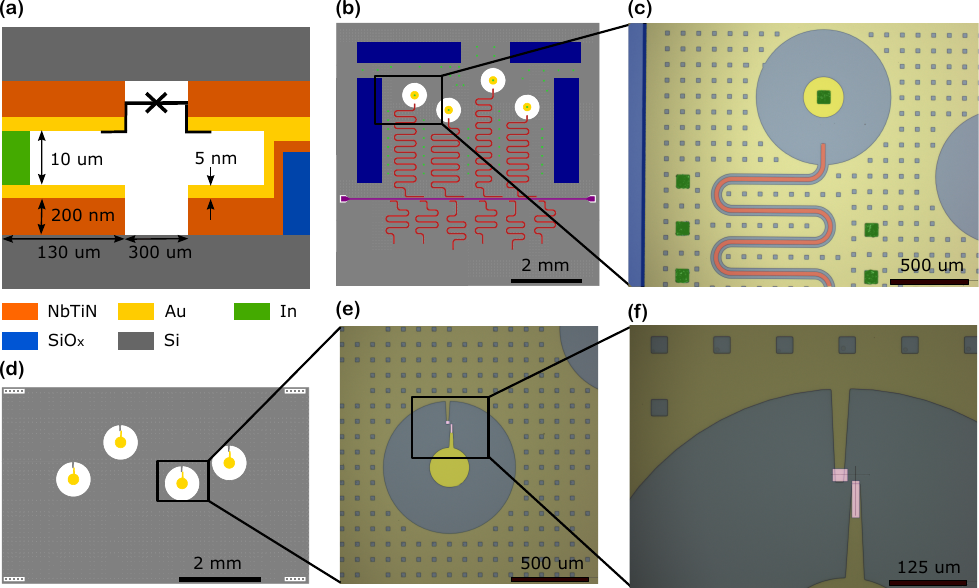}
    \caption{\textbf{Flip-transmon architecture:} (a) Cross section of a flip-transmon. The left side is the charge island, which contains an indium bump (green) maintaining the superconducting galvanic connection between the top and bottom islands. The right side is the ground, which has a SiO${}_x$ hardstop to ensure \qty{10}{\um} interchip distance. (b) Overview of the bottom chip design, which contains four charge islands (yellow), four readout resonators and 6 flag resonators (red), one transmission line (purple), four SiO${}_x$ hardstops (blue), and indium bumps across the ground plane and charge islands (green). (c) Optical image of the bottom chip showing a close-up of the annotated region in (b). (d) Overview of the top chip design, which contains 4 charge islands (yellow), each with a Josephson junction short to the ground (gray). (e) Optical image showing a close-up of a transmon island on the top chip. (f) Magnified image of the Al/AlO${}_x$/Al Josephson junction in (e) deposited on the Au film with Al patches covered on top.}
    \label{device}
\end{figure*}

In this work, we develop a three-dimensional transmon architecture designed for future hybrid semiconductor-superconductor quantum devices. The shunt capacitor is distributed across two bonded substrates such that the electric field is shared equally between the chips while maintaining low energy participation at the indium-bump interface.

Using this platform, we investigate the compatibility of electroplated indium interconnects with high-coherence superconducting circuits. We fabricate and characterize flip-chip devices in which electroplated indium bumps provide both galvanic connections within the qubit capacitor and ground-plane connections between chips. In addition, we perform a systematic study using coplanar-waveguide resonators to identify losses associated with the electroplating process. 

\section{Design and Fabrication}

To establish a platform for qubits that are hybrids of two different substrates, we design a 3D-integrated transmon qubit, see Fig.~\ref{device}. Although the substrates in principle could be of different materials, we will in this work focus on silicon substrates only. In contrast to conventionally coplanar transmon qubits, we integrate an indium bump directly into the transmon island, such that the capacitor of the transmon is galvanically connected across both top and bottom chip, see Fig.~\ref{device}(a). The process flow for the fabrication of this device is as follows: First, we deposit 10~$\mu$m of silicon-oxide (SiO${}_x$) on a high-resistivity wafer~\cite{Rebecca2026}. Using a buffered oxide etch, we define four rectangles of SiO${}_x$ which will act as hard-stops for the flip-chip process~\cite{8993515, Norris2024}, see Fig.~\ref{device}(b). Next, the full wafer, including the hard-stops, are covered by a \qty{200}{\nm} niobium-titanium-nitride (NbTiN). To ensure good electrical and mechanical connection between the base layer and the indium bumps during electroplating, the NbTiN layer is further covered with \qty{5}{\nm} of gold (Au). Note that Au is deposited ex-situ in a different deposition system than NbTiN in contrast to recent work on in-situ encapsulated superconducting circuits~\cite{Makita2026, Bal2024, PhysRevLett.134.097001}. The chip contains four transmon islands, see Fig.~\ref{device}(b) and (c), each with an 80 by \qty{80}{\um} indium bump in the center. Note that the indium bumps are squares in our design as we found a higher yield of the indium process with square bumps as compared to circular bumps. All four transmon islands are capacitively coupled to a $\lambda/2$ coplanar waveguide (CPW) readout resonator each of which are capacitively coupled to the feedline. Additional indium bumps are distributed across the ground plane to ground the top chip. Prior to the flip-chip bonding, the chip is cleaned with oxygen plasma to remove any excess resist residues on the indium bumps, see also Appendix~\ref{fab} for additional details of the fabrication process.

The top chip is also fabricated using NbTiN on a silicon substrate which is subsequently covered with Au. In contrast to the bottom chip, there are no hard-stops on the top chips. Instead, the top chip only contains the four transmon islands as well as the Josephson junctions of each transmon, see Fig.~\ref{device}(d), (e) and (f). The Josephson junctions are Al/AlO${}_x$/Al Manhattan-style junctions fabricated with a lift-off process. Note that the Josephson junctions are deposited directly on top of the gold layer and, to improve the electrical contact between the gold and the aluminum junction leads, a large aluminum patch is added to increase the overlap area between the gold and aluminum layers~\cite{10.1063/1.4993577}.

\begin{figure}[t]
    \centering
    \includegraphics[width=0.48\textwidth]{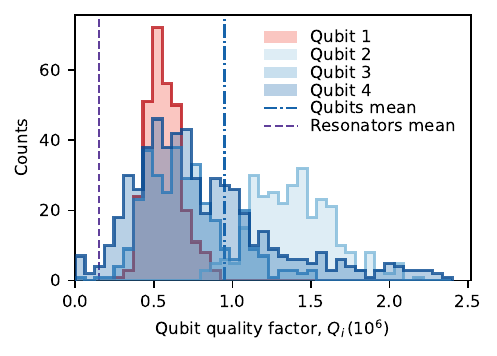}
    \caption{\textbf{Measurement of flip-transmons.} Distribution of the measured internal quality factors for four flip-transmons over a period of approximately 20 hours. Qubit 1 belongs to a different chip than Qubits 2, 3, and 4. The two chips have identical designs and were fabricated in the same processing run. The blue dashed-dot line indicates the mean internal quality factor of the four qubits. The purple dashed line indicates the mean internal quality factor of the resonators located on the same chip as Qubits 2, 3, and 4, corresponding to the lower-half structures shown in Fig.~\ref{device}(b).}
    \label{Flipmon Qi}
\end{figure}

\section{Loss characterization}

We measure qubit lifetimes of 4 different transmon qubits all fabricated as described above, see the results in Table~\ref{table:t1}, in order to characterize the quality of the bump bonds. The transmon qubits are from two different chips such that transmon 1 is from one chip while transmons 2, 3, and 4 belong to the second chip. The two chips have identical designs and were fabricated and cooled down together. We measured the $T_1$ lifetime of each qubit, over a timespan of around 20 hours, see the histogram of the recorded quality factors, $Q_i = \omega T_1$ where $\omega$ is the qubit angular frequency, in Fig.~\ref{Flipmon Qi}. The measured qubit quality factors are distributed mostly between $Q_i = 0.5\times10^6$ and $Q_i = 1.5\times10^6$, with variations both between qubits and over time. These fluctuations are consistent with the temporal variations commonly observed in superconducting qubits, for example, due to coupling to microscopic defects or other time-dependent loss channels~\cite{PhysRevLett.121.090502, PhysRevB.92.035442}. The Purcell limits for all qubits are above 10 ms, which is orders of magnitude higher than measured $T_1$. Importantly, all four qubits show quality factors of the same order of magnitude, despite being fabricated on two different chips, indicating that the process is reproducible across the two chips investigated here. 

To identify whether the measured qubit lifetimes are limited by the indium bump bonds or by other loss channels, we compare the qubit quality factors to the internal quality factors of the CPW flag resonators. As shown in Fig.~\ref{device}, the resonators are on the same chip and, hence, fabricated from the same baselayer and undergo the same processing steps as the transmon islands, including the electroplating of the indium bumps prior to the definition of the microwave circuit. However, in contrast to the transmon mode, the resonators do not have an indium bump on the center strip. Moreover, the resonator gap between the transmission line and ground is \qty{17.5}{\um}, much smaller than the transmon gap of \qty{300}{\um}, and therefore we expect the electric field of the resonator to be more strongly concentrated near the surface layers. With these considerations in mind, the mean of the measured resonator quality factors are shown together with the qubit quality factors in Fig.~\ref{Flipmon Qi}. We find that the resonator quality factors are systematically lower than the qubit quality factors. This observation suggests that the dominant loss mechanism is not the resistive or surface losses associated with the indium bump bond itself. Instead, the lower resonator quality factors suggest that dielectric loss at the interfaces is likely the limiting mechanism, since the resonators have larger surface participation ratios~\cite{10.1063/1.3637047, 10.1063/1.4934486}.

\begin{table}
\begin{tabular*}{\columnwidth}{@{\extracolsep{\fill}}lccc@{}}
\toprule
\toprule
           & $\omega/2\pi$ (GHz) & $T_1$ $(\mu s)$  & $Q_{i}$ $(10^6)$ \\ \midrule
Transmon 1$^*$ & 2.85      & $31.0^{+0.7}_{-0.7}$   & $0.56^{+0.01}_{-0.01}$                    \\
Transmon 2 & 2.98      & $74.8^{+1.8}_{-1.7}$ & $1.40^{+0.03}_{-0.03}$                     \\
Transmon 3 & 2.74      & $42.3^{+2.1}_{-2.0}$  & $0.73^{+0.04}_{-0.03}$                      \\
Transmon 4 & 2.75      & $60.4^{+7.4}_{-6.4}$  & $1.05^{+0.13}_{-0.11}$                  \\ \bottomrule\bottomrule
\end{tabular*}
\caption{\textbf{Flip-transmon relaxation times.} Relaxation time $T_1$ sampled over a duration of around 20 hours. The internal quality factor $Q_{i}$ is calculated by $Q_{i}$ = $\omega\, T_1$, where $\omega$ is the qubit angular frequency. The $^*$ indicates measurements on a different chip than the other qubits. The error bars indicate a 68\% confidence interval of the mean extracted from a bootstrapping estimate.} \label{table:t1}
\end{table}

\begin{figure*}
    \centering
    \includegraphics[width=1\textwidth]{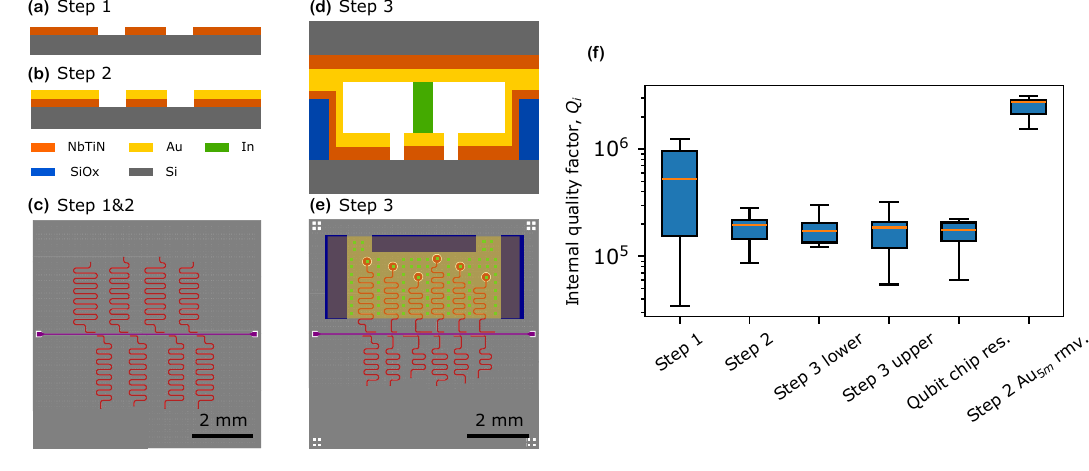}
    \caption{\textbf{Resonator measurements for loss analysis.}  (a) Cross section of resonators fabricated through Step 1, consisting of \qty{200}{\nm} NbTiN deposited on a high-resistivity silicon substrate and patterned by RIE. (b) Cross section of resonators fabricated through Step 2, which includes an additional \qty{5}{\nm} ex-situ Au film deposited on top of the NbTiN. (c) Layout of the CPW resonators used for the Step 1 and Step 2 measurements. (d) Cross section of resonators fabricated through Step 3. Additional processing includes deposition and patterning of SiO${}_x$ hard stops, electroplated indium bumps, and bonding to a blank top chip. (e) Layout of the resonators used for the Step 3 measurements. The lower half contains resonators that experience only the SiO${}_x$ processing steps, while the upper half includes electroplated indium bumps connected to a flipped top chip. (f) Measured internal quality factors in the low photon regime for resonators fabricated through the different process steps. Columns 1–2 correspond to Steps 1 and 2, columns 3–4 correspond to the structures shown in (e), column 5 shows resonators fabricated on the flip-transmon chips, and the final column shows Step 2 resonators after removal of the Au film. The orange lines indicate the median quality factors, the boxes indicate the first and third quartile of the data while the whiskers indicate the minimum and maximum data points.}
    \label{figure3}
\end{figure*}

To further investigate the origin of the dominating loss mechanism, we fabricate sets of superconducting resonators through the different process steps used for the flip-transmon devices, as summarized in Fig.~\ref{figure3}. In Step~1, the resonators are fabricated from a \qty{200}{\nm} NbTiN film deposited on a high-resistivity silicon substrate and patterned by reactive-ion etching (RIE), see Fig.~\ref{figure3}(a) and (c). In Step~2, we add the same \qty{5}{\nm} Au film used in the flip-transmon process on top of the NbTiN film, see Fig.~\ref{figure3}(b) and (c). We later also remove the Au film from a subset of Step~2 resonators using the same Au wet etch used to define the transmon islands and resonators on the flip-transmon chips. Finally, Step~3 includes the full additional processing required for the 3D-integrated devices, including the initial deposition and patterning of the SiO${}_x$ hard stops, electroplating of the indium bumps, and bonding to a blank top chip, see Fig.~\ref{figure3}(d) and (e). We characterize the internal quality factor $Q_i$ of all resonators in the low photon regime (between 10 and 100 photons), such that the measurement is sensitive to the loss mechanisms relevant for superconducting qubits, see Fig.~\ref{figure3}(f). In total, we measured two chips for each step meaning between 12 and 16 resonators per step for a total of 75 resonators. The Step~1 resonators show quality factors in the range expected for patterned NbTiN resonators on silicon~\cite{10.1063/1.3458705, PhysRevApplied.11.064053}, while the addition of the thin Au film in Step~2 leads to a clear reduction of $Q_i$. Recent works have demonstrated that Au-encapsulated superconducting resonators may improve the quality factors when deposition in-situ with the base layer~\cite{PhysRevLett.134.097001}. However, we emphasize that in our work, Au is deposited in a separate deposition system. In Step~3, we also include $\lambda/4$-resonators that are shorted to ground through an indium bump that connects to the ground plane of the bare top chip. The resonators fabricated through Step~3, including the conventional resonators (lower part of the chip) and resonators shorted to ground with an indium bump (upper part of the chip), show quality factors comparable to the Au-coated Step~2 resonators. This correspondence between the resonator quality factors indicates that the electroplating process does not cause additional resistive losses at this level, in contrast to prior work on evaporated indium bumps~\cite{Rosenberg2017, granel20263dintegrationhybridquantum}. The resonators on the flip-transmon devices, as discussed above, also exhibit quality factors comparable to those of the Step~2 and Step~3 resonators, see Fig.~\ref{figure3}(f). After removing the Au film using a selective Au etch from a Step~2 chip, the resonator quality factors increase substantially, reaching values above $10^6$. This improvement indicates that the Au-covered metal-air interface is the dominant limiting interface in the Step~2 devices. We therefore conclude that Au deposition introduces losses at or near the exposed metal surface, most likely through impurities in the metal. On the other hand, these results also indicate that the Au film effectively protects the NbTiN surface since the resonators after Au removal exhibits higher quality factors than the bare NbTiN resonators.

\begin{figure}[t]
    \centering
    \includegraphics[width=0.48\textwidth]{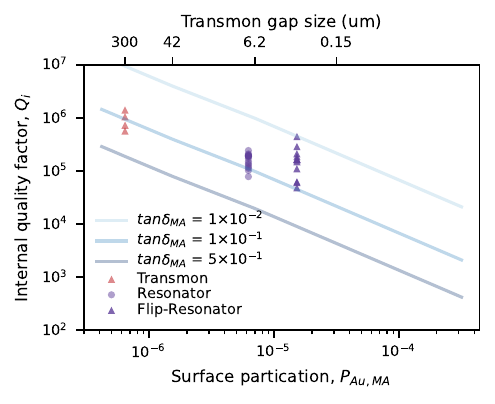}
    \caption{\textbf{Loss analysis.} Internal quality factor as a function of the simulated surface participation ratio of the metal-air (MA) interface where there is gold encapsulation for the Au-capped resonators from Step 2, the bump-bond grounded resonators from Step 3 that are covered with the top chip, and for flip-transmons in Fig.~\ref{device}.  The solid lines show the simulated relationship between $Q_i$ and the MA surface participation ratio for a fixed-$E_c$ flip-transmon with varying island diameter and island-to-ground gap, with the gap indicated on the upper x-axis.}
    \label{loss_analysis}
\end{figure}

To investigate to what extent the losses can be attributed to the Au metal-air (MA) interface, we simulate the corresponding surface energy participation ratio for the devices measured in Figs.~\ref{Flipmon Qi} and \ref{figure3}, see Fig.~\ref{loss_analysis}, which shows the measured internal quality factor as a function of the simulated Au MA participation ratio for the Au-capped resonators from Step~2, the bump-bonded resonators from Step~3, and the flip-transmon devices. The simulations were performed with ANSYS Maxwell 2D, see Appendix~\ref{EprSim}. By taking advantage of the symmetry of the resonator and flip-transmon geometries, and the implication of Fig.~\ref{figure3}(f) that the Au MA interface is the dominant loss source, we simplify the simulations to a two-dimensional cross-sectional model that includes only the metal structures, the MA interfaces and the silicon substrates. The transmons have the smallest simulated MA participation, $P_{\mathrm{Au,MA}}\sim 10^{-6}$, as expected since they have the largest gap, and show quality factors on the order of $10^6$. The Step~2 resonators have a larger participation, $P_{\mathrm{Au,MA}}\sim 10^{-5}$, agreeing with the lower quality factors, mostly in the range $Q_i\sim3\times10^5$. The flip-resonators have comparable or slightly larger MA participation than the resonators, but their quality factors extend to somewhat higher values, with a spread that reflects the device-to-device and temporal variations discussed above. Taken together, the three device classes show the expected inverse correlation between $Q_i$ and $P_{\mathrm{Au,MA}}$, as expected for a loss channel dominated by the surface interface. To further highlight the expected dependence on the Au MA interface, we also calculate the quality factor for a range of assumed loss tangents for this interface, see the solid lines in Fig.~\ref{loss_analysis}. Additionally, we performed flip-transmon simulations with the charging energy $E_c$ fixed while the island diameter and island-to-ground spacing are varied, giving the expected range of Au MA participation for transmon geometries with similar anharmonicity. The upper x-axis shows the equivalent gap between the charge island and the ground plane corresponding the surface participation values on the lower x-axis. It shows that a transmon gap footprint of around \qty{5}{\um} results in surface participation comparable to that of the two resonator designs. The measured data for all three device types are broadly consistent with an effective Au MA loss tangent between $\tan\delta_{\mathrm{MA}}\sim 5\times10^{-2}$ and $10^{-1}$. This agreement supports the conclusion that the Au-covered metal-air interface is the dominant loss channel in the present devices. In particular, the data are consistent with the resonator process study in Fig.~\ref{figure3}, where removing the Au film improves the resonator quality factors, indicating that the exposed Au surface limits the measured coherence rather than other interfaces such as the bump bonds. 

\section{Discussion and Outlook}
We have established a three-dimensional-integrated transmon qubit which could provide a platform for future hybrid semiconductor-superconductor qubits~\cite{PhysRevB.109.075101, PRXQuantum.6.010308}. In particular, we investigated electroplated indium bumps for flip-chip integration of superconducting quantum circuits. In contrast to conventional flip-chip test structures, the indium bump is incorporated directly on the transmon island. The resulting electric-field participation ratio is approximately 49\% in each substrate and nearly zero at the indium bump surface, as also discussed in Appendix~\ref{app:sim}. This architecture may be particularly advantageous for superconducting-semiconductor integration, since it allows for the engineering of the electric-field participation in the top substrate and, thus, make the qubit  tolerate the presence of a relatively lossy semiconductor substrate while suppressing sensitivity to losses associated with the indium bump interface. We measure qubit quality factors on the order of $10^6$, demonstrating that electroplated indium bumps can provide superconducting galvanic interconnects compatible with superconducting qubits. At the same time, the resonators on the same chip have systematically lower quality factors than the qubits, indicating that the measured coherence is not limited primarily by the indium bump itself, but rather by dielectric losses at the interfaces.

To identify the dominant loss channel, we fabricated and measured resonators at different stages of the fabrication process and compared the measured quality factors to simulations of the Au metal-air surface participation. The resonator measurements show that adding a thin Au film to the NbTiN film reduces the internal quality factor, while removing the Au film significantly improves it. Together with the energy participation analysis, this indicates that the Au metal-air interface is the dominant limiting interface in the present devices. These results suggest that future improvements should focus on reducing the participation and loss tangent of this interface, for example by replacing the ex-situ Au film using in-situ deposition. Such improvements would make electroplated indium bumps a promising route towards scalable flip-chip architectures for superconducting circuits and hybrid semiconductor-superconductor quantum devices.

\section*{Data availability}
All experimental data are available through \cite{Data_repository} and the code used for data processing is available through \cite{GitHub_repository}.

\section*{Acknowledgments}
Y.S. designed the resonator, flip-resonator, and flip-transmon devices. Y.S. and R.G., B.K., D.J., S.S., and L.J.S. fabricated the devices. Y.S., M.F.S.Z, E.Y.H, and N.Z. measured the experimental data. Y.S. analyzed the data. Y.S. wrote the manuscript with input from all coauthors, and S.G. and C.K.A. supervised the work. The authors thank the following for input and preliminary work that helped enable this work: P. Viswanathan, K. Czerniak, J. Winkelhorst, and E. Pot.
This research was co-funded by the Dutch Research Council (NWO), the European Innovation Council Pathfinder Grant No. 101115315 (QuKiT), NWO Talent Programme Vidi Science domain (VI.Vidi.233.031) and by Holland High Tech (TKI) project 00PPS334. 

\appendix
\section{Fabrication}
\label{fab}

\begin{figure*}
    \centering
    \includegraphics[width=1\textwidth]{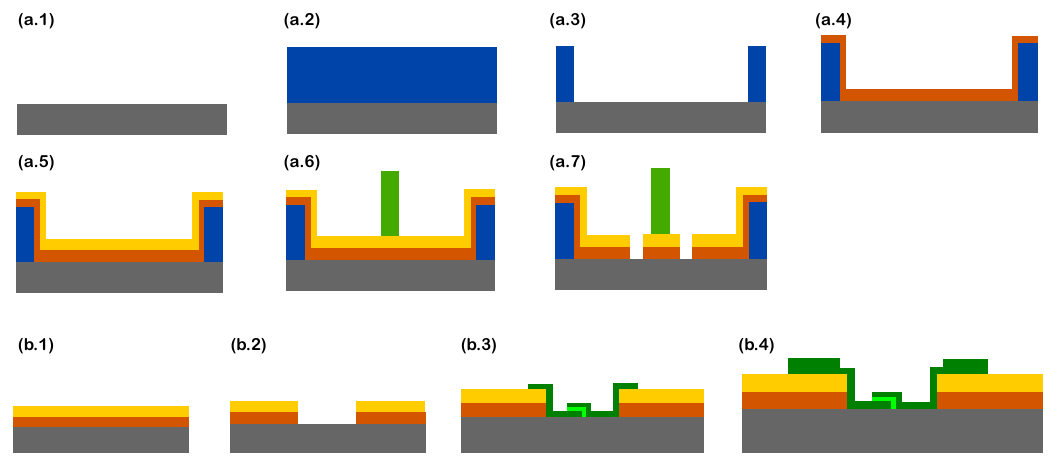}
    \caption{\textbf{ Fabrication flow of the flip-transmon device.}  (a.1)–(a.7) Bottom chip fabrication process. (a.1) Starting high-resistivity Si wafer and cleaning steps. (a.2) Deposition of the SiO${}_x$ hard stops. (a.3) Patterning of the SiO${}_x$ hard stops. (a.4) NbTiN deposition. (a.5) Ex-situ Au deposition. (a.6) Indium bump electroplating. (a.7) NbTiN and Au etching. (b.1)–(b.4) Top chip fabrication process. (b.1) NbTiN and Au deposition on a cleaned Si wafer. (b.2) Patterned Au and NbTiN layers. (b.3) Fabrication of (Al/AlO${}_x$/Al) Josephson junctions. (b.4) Aluminum patches.} \label{fig:fab}
\end{figure*}

The superconducting qubit fabrication process used here is based on the process reported in Ref.~\cite{yxf3-jtx5} with a few extra steps that we will discuss here and which is illustrated in Fig.~\ref{fig:fab}. For the bottom chip, shown in Fig.~\ref{fig:fab}(a.1), the process starts with a high-resistivity silicon wafer (>20k$\Omega\cdot$cm) with a (100) orientation, sourced from Topsil. The wafer is first immersed in acetone for 5 minutes with sonication to remove organic residues. It is then transferred to isopropyl alcohol for 60 seconds to remove any remaining acetone. An anisotropic oxygen plasma descum is subsequently performed using 20 sccm O${}_2$ at 0.1 mbar and 400 W for 30 seconds to further clean the surface. The native oxide is removed using buffered oxide etch (BOE) for 5 minutes, followed by a 60-second rinse in deionized water. Immediately after the BOE step, the wafer is loaded into Plasmalab 80 Plus (Oxford Instruments) for Plasma-enhanced chemical vapor deposition (PECVD). As shown in Fig.~\ref{fig:fab}(a.2), a $10~\mu\mathrm{m}$ thick SiO${}_x$ layer is deposited at a rate of 70 nm/min~\cite{Rebecca2026}. To pattern the hard stops, AZ nLOF 2020 is used as the photoresist. The resist is spin-coated at 4000 rpm and soft-baked at $100~^\circ\mathrm{C}$ for 90 seconds. Photolithography for the hard-stop patterns is performed using a Heidelberg Instruments laserwriter $\mu$MLA. After developing the resist, an anisotropic oxygen plasma descum is then carried out using 20 sccm $O_2$ at 0.1 mbar and \qty{300}{\W} for 30 seconds to remove any organic residues. Finally, the wafer is immersed in BOE for 22 minutes with sonication to etch the SiO${}_x$ layer and define the hard stops, as shown in Fig.~\ref{fig:fab}(a.3). After the etching process, the remaining resist is removed by immersing the wafer in nitric acid for 5 minutes, followed by a 1 minute rinse in deionized water.

After patterning the SiO${}_x$ hard stops, \qty{200}{\nm} of NbTiN is deposited as shown in Fig.~\ref{fig:fab}(a.4), see also Ref.~\cite{yxf3-jtx5}. The vacuum is then broken, and the wafer is transferred to a different evaporation system (supplied by AJA International) for gold deposition without any surface cleaning between the two processes. The transfer from one system to the other is performed in less than 5~minutes. A \qty{5}{\nm} thick Au film is deposited at a rate of 0.1 nm/s, see Fig.~\ref{fig:fab}(a.5).

With the Au film serving as an electrically conductive seed layer at room temperature, the indium bump electroplating process is carried out. First, AZ40XT photoresist is spin-coated at \qty{3000}{rpm} and soft baked at \qty{120}{^\circ\mathrm{C}} for 7 minutes. Photolithography is then performed to define the indium bump pattern.  Indium electroplating is performed using a Smart Cell 1000W system from YAMAMOTO-MS. A ready-to-use indium sulfamate supplied by Indium Corporation is used as the plating bath. A pure indium bar, also provided by Indium Corporation, serves as the anode, while an in-house-made sample holder is used as the cathode. A unipolar pulse sequence with a current of \qty{5}{mA} is applied for around 10 minutes aiming for a bump height of \qty{20}{\um}. The whole electroplating process is done under room temperature. Finally, the photoresist is stripped in NMP at \qty{80}{^\circ\mathrm{C}} for 2 hours, see Fig.~\ref{fig:fab}(a.6).

After the indium bump electroplating process, the Au and NbTiN layers are patterned using photolithography with AZ10XT photoresist. The exposed resist is developed in a solution of AZ400K and deionized water with a volume ratio of 1:4. Before etching, the sample undergoes an oxygen plasma descum for 2 minutes to remove any residual photoresist remaining on the surface. The Au and NbTiN layers are etched in two separate processes using the same photoresist mask. The Au film is first removed using a standard potassium-iodide-based gold etchant supplied by Sigma-Aldrich. The chip is immersed in the etchant for 5 seconds and subsequently rinsed in deionized water for 60 seconds. The NbTiN layer is etched using reactive-ion etching with a two-step recipe. In the first step, an SF${}_6$/O${}_2$ gas mixture with a flow ratio of 13.5:4 is used at \qty{70}{W} and a pressure of \qty{0.1}{mbar}. The etch is continued until a signal drop is detected by the endpoint detection system. In the second step, an SF${}_6$ gas mixture with a flow ratio of 4:16 is used at \qty{50}{W} and a pressure of \qty{0.08}{mbar}. After the etching process, the photoresist is removed by immersing the sample in NMP for 2 hours. Before wire bonding the device to a printed circuit board (PCB), an oxygen plasma descum is performed to remove any remaining organic residues. The completed bottom chip is shown in Fig.~\ref{fig:fab}(a.7).

For the top chip, a high-resistivity silicon wafer is first cleaned in nitric acid for 7 minutes, followed by a 6 minute dip in a (40\%) HF solution. The wafer then undergoes the same NbTiN and Au deposition processes used for the bottom chip, as shown in Fig.~\ref{fig:fab}(b.1), followed by the same Au and NbTiN etching procedures, as shown in Fig.~\ref{fig:fab}(b.2). The aluminum-based Josephson junctions, Fig.~\ref{fig:fab}(b.3), and aluminum patches, Fig.~\ref{fig:fab}(b.4), are subsequently fabricated using the same process described in Ref.~\cite{yxf3-jtx5}. Finally, we bond the top and bottom chip using a Tresky Die bonder T300 under a force of \qty{196}{N}.

\section{Experimental setup}
\begin{figure}
    \centering
    \includegraphics[width=0.48\textwidth]{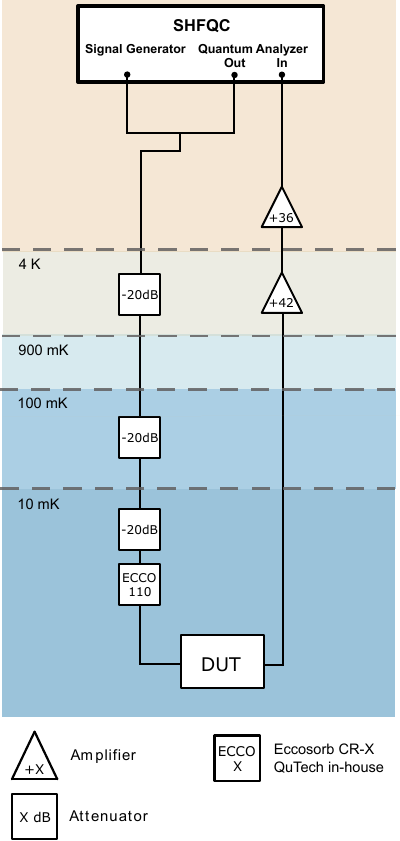}
    \caption{\textbf{Wiring diagram of the experimental setup.}}
    \label{fridge}
\end{figure}

The experimental setup used for device characterization is shown in Fig.~\ref{fridge}. All experiments are performed in a Bluefors LD400 dilution refrigerator at a base temperature around \qty{10}{mK}. The sample is protected from thermal and electromagnetic radiation by a copper can and two mu-metal shields. The readout and drive pulses are generated and analyzed by a Zurich Instruments SHFQC as part of an OrangeRack delivered by \textit{Orange Quantum Systems}.  All signals pass through a series of attenuators, filters, and in-housemade Eccosorb infrared filters with a total attenuation of \qty{75}{dB}. The output signal is amplified by a cryogenic HEMT amplifier (LNF-LNC4-8C) and a room-temperature  amplifier (AFS4-00100600-13-10P-4).

\section{Magnetic field dependence}
\begin{figure}
    \centering
    \includegraphics[width=0.48\textwidth]{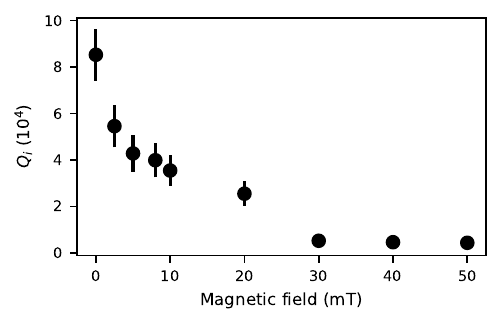}
    \caption{\textbf{Characterization of interconnection.} Internal quality factor of step 3 upper resonators as a function of in-plane magnetic field measured at $\langle n_{ph} \rangle \simeq 10^7$. The error bars indicate a 68\% confidence interval of the mean extracted from a bootstrapping estimate.}
    \label{indium bump}
\end{figure}

To assert if the indium bumps provide good and low-resistivity electrical connection between the two chips, we measured quarter-wavelength coplanar-waveguide resonators that have an indium bump located at the current node, providing a short to the ground plane on the top chip. In addition to the measurements discussed in the main text where the sample is enclosed in magnetic shielding, we also measure the device in a 6-1-1 vector magnet. As the in-plane magnetic field is increased, the internal quality factors of the resonators decrease and eventually converge, see Fig.~\ref{indium bump}, consistent with the behavior reported in Ref.~\cite{granel20263dintegrationhybridquantum}. The drop in quality factor can be readily understood from the closing of the superconducting gap in the indium bump before the bumps transition to a normal metal. Also note that the device is, for these measurements, not encapsulated by the magnetic shielding used in the main text, thus, the initial quality factors at zero field are lower than the corresponding quality factors in the main text.

\label{app:sim}
\begin{figure*}
    \centering
    \includegraphics[width=1\textwidth]{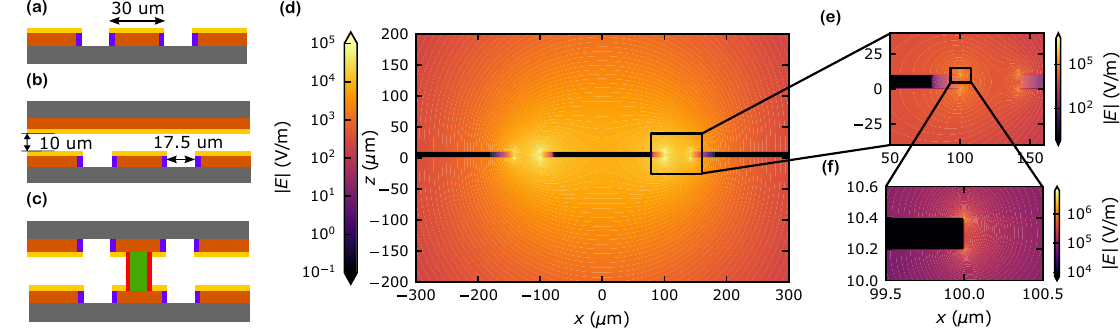}
    \caption{\textbf{ EPR simulation}  (a) Schematic of a simulated CPW resonator. (b) Schematic of a simulated flip-CPW resonator. (c)  Schematic of a simulated flip-transmon. (d) The simulated electric field distribution of a flip-transmon. (e) Magnified image of the electric field distribution around the indium bump and transmon gap. (f) Magnified image of the electric field distribution on the edge of the charge island.}
    \label{Maxwell}
\end{figure*}

\section{Simulations of the surface energy participation ratios}
\label{EprSim}

To calculate the energy participation ratio (EPR) at each dielectric element, we perform finite-element simulations of the resonator and transmon geometries. For the purpose of these simulations, all materials are assumed to be lossless. The base layer is modeled as a \qty{200}{\nm} perfect conductor, while the indium bump is modeled as a cylindrical perfect conductor with a radius of \qty{100}{\um}. Based on the resonator measurements presented in Fig.~\ref{figure3}(f), the metal-air (MA) interface is inferred to be the dominant loss source. Therefore, the only interface included in the simulation, in addition to the bulk substrates, is the MA interface. The metal-substrate (MS) and substrate-air (SA) interfaces are neglected. To distinguish between different MA interfaces, distinct \qty{5}{\nm} thick dielectric layers are assigned to the Au surface, the etched sidewalls, and the indium bump surface.
 
Figures~\ref{Maxwell}(a)-(c) show the simulation models of the CPW resonator, flip-resonator, and flip-transmon used in the main text. The gray regions represent the silicon substrate, the orange regions represent the perfectly conducting base layer, and the green regions represent the perfectly conducting indium bumps. The yellow, purple, and red layers correspond to the Au MA interface, the etched sidewall MA interface, and the indium bump MA interface, respectively. All MA interface layers are modeled as \qty{5}{\nm} thick dielectric layers with an effective relative permittivity of 10.

The CPW resonator and flip-resonator are simulated using ANSYS Maxwell 2D in the Cartesian XY geometry mode. Although the fabricated flip-resonator contains an indium bump on the center conductor, it is omitted from the simulation because the bump is located at the voltage antinode, where its effect on the electric field distribution is negligible. The flip-transmon is simulated using the cylindrical-about-Z geometry mode to take advantage of its rotational symmetry. As shown in Fig.~\ref{Maxwell}(d), the electric field for the flip-transmon is primarily confined to the transmon gap and the silicon substrates. Fig.~\ref{Maxwell}(e) provides a magnified view of one side of the transmon and the indium bump, illustrating that the electric field decreases rapidly away from the island and ground-plane edges. Fig.~\ref{Maxwell}(f) further zooms in on the edge of the top island, where the electric field is concentrated at the corners, as expected.

We calculate the EPR for the surface as
\begin{equation}
p_{\mathrm{MA}}
=
\frac{\int_{\mathrm{MA}}\varepsilon_r |\mathbf{E}|^2\, dV}
{\int_{\mathrm{all}}\varepsilon_{all}\cdot|\mathbf{E}|^2\, dV}
\label{eq:MA}
\end{equation}
and for the bulk as
\begin{equation}
p_{\mathrm{bulk}}
=
\frac{\int_{\mathrm{bulk}}\varepsilon_r |\mathbf{E}|^2\, dV}
{\int_{\mathrm{all}}\varepsilon_{all}\cdot|\mathbf{E}|^2\, dV}.
\label{eq:bulk}
\end{equation}
The results are summarized in Table~\ref{table:EPR} and, as expected, the Au MA participation ratio increases as the capacitor gap decreases, comparing the CPW resonator, flip-resonator, and flip-transmon~\cite{10.1063/1.4934486}. The Au MA participation further increases in the flip-resonator because of the additional gold-coated top ground plane. For the flip-transmon, 98.6\% of the electric field energy is stored in the silicon substrates, with the participation shared equally between the top and bottom chips.

\begin{table}
\begin{tabular*}{\columnwidth}{@{\extracolsep{\fill}}lccc@{}}
\toprule
\toprule
           & resonator & flip-resonator  & flip-transmon \\ \midrule
Au MA ($\times10^{-4}$\%) & 6.2      & 15.2   & 1.5 \\
bottom substrate (\%) & 92      & 80.9 & 49.3                               \\ 
top substrate (\%)& --      & -- & 49.3                               \\ 
\bottomrule\bottomrule
\end{tabular*}
\caption{\textbf{EPR across MA and substrates } Simulated energy participation ratio (EPR) of the Au metal-air (MA) interface and the silicon substrates for the CPW resonator, flip-resonator, and flip-transmon.} \label{table:EPR}
\end{table}

\newpage

\bibliography{reference}

\end{document}